# IMPACT OF FASTENERS ON THE RADAR CROSS-SECTION PERFORMANCE OF RADAR ABSORBING AIR INTAKE DUCT

Vijay Kumar Sutrakar[1] and Anjana P K[1]

[1]Aeronautical Development Establishment, DRDO
Bangalore, India

**ABSTRACT**

*An aircraft consists of various cavities including air intake ducts, cockpit, radome, inlet and exhaust of heat exchangers, passage for engine bay/other bay cooling etc. These cavities are prime radar cross-section (RCS) contributors of aircraft, especially at frontal sector of the aircraft. These cavities are of different cross-sectional area as well as different length depending on the applications on aircraft. The major such cavity is air intake duct and it contributes significantly to frontal sector RCS of an aircraft. The RCS reductions of air intake duct is very important to achieve a low RCS (or stealthy) aircraft configuration. In general, radar absorbing materials (RAM) are getting utilized for RCS reduction of air intake duct. It can also be noticed that a large number of fasteners are used for integration of air intake duct with the aircraft structures. The installation of fasteners on RAS may lead to degradation of RCS performance of air intake. However, no such studies are reported in the literature on the impact of rivets on the RCS performance of RAS air intake duct. In this paper, radar absorbing material of thickness 6.25 mm is designed which givens more than -10 dB reflection loss from 4 to 18GHz of frequencies. Next, the effect of rivet installation on these RAS is carried out using three different rivet configurations. In all the three cases the percentage of rivets heads area to RAS area are different, i.e. (I) 9.8%, (II) 4.4%, and (III) 2.5%. The RCS performance of RAS is evaluated for duct of different lengths from 1 to 18GHz of frequencies. In order to see the RCS performance, five different air intake cases are considered The RCS performance with increase in percentage surface area of rivet heads to RAS is reported in details. At the last, an open-source aircraft CAD model is considered and the RCS performance of RAS air intake with and without rivets is evaluated.*

## 1. INTRODUCTION

Stealth technology is an integral component of modern military defence systems, significantly enhancing the survivability of platforms by reducing their radar cross section (RCS). Air intake ducts, crucial for aerodynamic and engine performance, are also substantial contributors to an aircraft's electromagnetic signature [1] – [4]. Achieving a balance between aerodynamic efficiency and stealth remains a significant engineering challenge [5] – [6]. An aircraft consists of various other cavities including cockpit, radome, inlet and exhaust of heat exchangers, passage for engine bay/other bay cooling etc. [7] – [8]. These cavities are of different cross-sectional area as well as different length depending on the applications on aircraft. These cavities are prime radar cross-section (RCS) contributors of aircraft, especially at frontal sector of the aircraft [8] – [10]. The major such cavity is air intake duct and it contributes significantly to the frontal sector RCS of an aircraft. The RCS reductions of air intake duct is very important to achieve a low RCS (or stealthy) aircraft configuration. The preferable air intake duct shape of stealth aircraft is serpentine in nature [11]. Next, step is to utilize radar absorbing materials (RAM) for further reduction of RCS of air intake duct. In general, radar absorbing structures (RAS) are primarily made of different classes of RAMs embedded in glass fabric reinforced plastic (GFRP) [12] – [14]. However, carbon fibre reinforced plastic (CFRP) or metals are preferred choice (instead of GFRP) for the structural design of air intake duct of conventional aircraft. Also, it can be seen that a large number of fasteners are used for integration of air intake duct with the aircraft structures. The installation of fasteners on RAS may lead to degradation of RCS performance of air intake.

Weak scattering sources are often overlooked in RCS evaluations but can become dominant contributors in low-RCS designs. As [15] explain, weak scatterers, which are negligible in conventional aircraft, significantly affect stealth platforms due to their reduced overall radar visibility. Precise measurement and analysis of these sources are critical for optimizing stealth performance. This insight underscores the importance of examining all structural elements, including fasteners, which are distributed across the surfaces of any aircraft. Fasteners may introduce significant RCS challenges for a stealth aircraft [16]. The analysis of the electromagnetic scattering of rivets on the conducting plane using Electric Field Integral Equation (EFIE) is reported and the RCS effect of rivets on a plate with incident angle is studied. A more emphasize was given on the numerical simulation technique in [17].

The scattering from round-head screws can account for a measurable portion of a target's total RCS is highlighted [16]. Their study proposed optimized screw designs to suppress scattering, demonstrating the importance of tailoring fastener geometry for stealth applications. Maintenance and damage detection are essential for long-term stealth performance. A dual modality detection system combining visual and microwave imaging to assess damage in radar-absorbing materials is proposed [18]. This methodology offers potential for evaluating fastener integrity and their impact on RCS over time. In summary, while significant advancements have been made in RCS reduction technologies, the impact of fasteners on radar-absorbing air intake ducts remains unreported.

In the present work, impact of rivets on the RCS performance of realistic sized RAS air intake duct is reported for the first time. Section 2 deals with the modelling and simulation details. Results and discussions are presented in section 3, followed by concluding remarks in section 4.

## 2. MODELLING AND SIMULATION DETAILS

In this work, firstly, EM modelling and simulation of radar absorbing materials is carried out. Jerusalem cross design as perfectly electrical conducting (PEC) material at cross with resistance of 100 ohms on four arms has been designed [13]. The RAM used in the present design consists of three layers. First and third layers consist of FR4 material with the dielectric constant of $\epsilon_r = 4.4$, and loss tangent of $tan\delta = 0.02$. Second layer consists of the Jerusalem cross unit cell based periodic structure design bonded with FR4 material of thickness 0.125mm. The total thickness of the sample is 6.25 mm. The reflection coefficients of RAM have been computed by assigning Floquet port with periodic (lateral) and PEC (bottom layer) boundary conditions using full wave simulation technique from 1 to 30 GHz of frequencies [19]. The proposed RAM shows more than -10 dB reflection loss from 4 to 18 GHz of frequencies.

Next, the effect of rivet installation on these RAS is carried out using three different rivet configurations, i.e. (a) rivet installed at the centre of unit cell (RAS C1), (b) rivet installed between the gap of two-unit cells (RAS C2), and (c) rivet installed at the centre of four-unit cells (RAS C3). All the three rivet configurations are shown in Figure 1.

Next, three cases with varying the percentage of rivets heads area to RAS surface area are studied, i.e. (a) 9.8%, (b) 4.4%, and (c) 2.5%. The rivet of 4.0mm diameter with countersunk of 90 degree is considered. Reflection coefficients of all these configurations have been computed using Floquet port analysis [19]. This study is carried out to understand the impact of conventional fastening on the RAS performance. However, a simplified case of rivet is considered in the paper. More detailed study will be an area of open research.

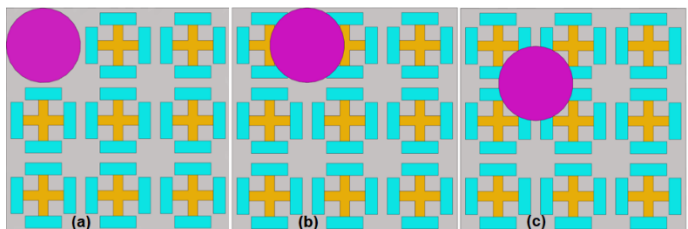

**FIGURE 1:** DIFFERENT RIVET CONFIGURATIONS (a) RAS C1, (b) RAS C2, AND (c) RAS C3

In order to evaluate the performance of these rivet configurations, initial studies are carried out on a flat panel sample of dimensions 300mmx300mm. The flat panel are simulated from 1 to 30GHz of frequencies with a step size of 1GHz for both horizontal and vertical polarizations for ϴ=0 degree and ϕ varying from -90 degree to +90 degree with a step of 0.5 degree.

Subsequently, the performance of rivets is evaluated on different circular air intake cavities, i.e. metallic intake, RAS intake without rivets, and nine cases of rivets with RAS as discussed above, are considered. The air intake with diameter of 25λ and length of 16.67λ, 33.33λ, 66.67λ, 133.34λ, 200λ, and 266.67λ corresponding to 10GHz is considered. The high frequency EM simulations using shooting and bouncing rays are used for cavity study with 20 number of bounces with 5 ray densities [19]. Simulation is carried out from 1 to 18GHz of frequencies with a step size of 1GHz for both horizontal and vertical polarizations for ϴ=0 degree and ϕ varying from 180 degree to 240 degree with a step of 0.5 degree. Median RCS at a given frequency is also calculated for varying ϕ from 180 degree to 240 degree for a given polarization.

At the last, an open-source aircraft CAD model is considered, as shown in Figure 2 [20] – [21]. The performance of RAS air intake with and without rivets is evaluated and results are compared with aircraft with non-RAS air intake. The high frequency EM simulations using shooting and bouncing rays are used for cavity study with 20 numbers of bounces with 5 ray densities [19]. Simulation is carried out from 1 to 18GHz of frequencies with a step size of 1GHz for both horizontal and vertical polarizations for ϴ=0 degree and ϕ varying from 180 degree to 270 degree with a step of 0.5 degree. Median RCS at a given frequency is also calculated for varying ϕ from 180 degree to 240 degree for a given polarization.

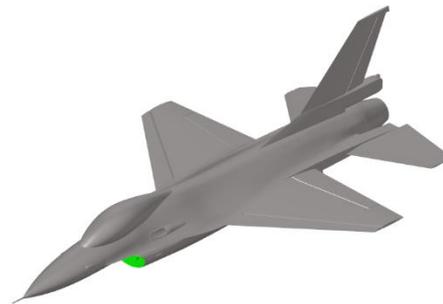

**FIGURE 2:** OPEN-SOURCE AIRCRAFT CAD MODEL [20]

## 3. RESULTS AND DISCUSSIONS
### 3.1 Floquet Port Analysis

The reflection coefficients of RAM have been computed by assigning Floquet port with periodic (lateral) and PEC (bottom layer) boundary conditions using full wave simulation technique from 1 to 30 GHz of frequencies [19]. The proposed RAM without rivet shows more than -10 dB reflection loss from 4 to 18 GHz of frequencies, as shown in Figure 3. Next, the effect of rivet installation on these RAS is carried out using three different rivet configurations (refer Section 2 for further details). Figure 3

(a) shows the reflection coefficients of different RAS with and without rivets (with rivet surface area of 9.8%). It can be seen that the different rivet locations are leading to different reflection coefficients with varying frequencies. For example, RAS C1 shows maximum of 10dB degradation of S-parameters at ~10GHz of frequency, as shown in Figure 3(a). RAS C2 also shows approximately 10dB degradation in reflection coefficient at ~13GHz of frequency. RAS C3 shows better performance compared to RAS C1 and RAS C2 for rivet surface area of 9.8%. However, the performance keeps varying with frequencies. Result shows that rivet surface area of 9.8% leads to significant degradation in reflection coefficient across the frequencies at normal incident angle.

Next, three cases with varying the percentage of rivets heads area to RAS surface area are studied, i.e. (a) 9.8%, (b) 4.4%, and (c) 2.5%. The rivet of 4.0mm diameter with countersunk of 90 degree is considered. Reflection coefficients of all these configurations have been computed using Floquet port analysis, as shown in Figures 3(a), 3(b), and 3(c) for rivet surface area of 9.8%, 4.4%, and 2.5%, respectively. Result shows that with reduction in rivet surface area to RAS surface area, the degradation in reflection coefficients reduces. Rivet surface area of 4.4% shows reflection coefficient degradation of maximum 5dB and rivet surface area of 2.5% shows maximum of 3dB (except null regions) at normal incident angle, as shown in Figures 3(b) and 3(c), respectively.

### 3.2 Effect of rivet configurations on flat RAS panels

In this section, mono-static RCS performance of flat panel RAS sample of dimensions 300mmx300mm is evaluated. The flat panel are simulated from 1 to 30GHz of frequencies with a step size of 1GHz for both horizontal and vertical polarizations for Θ=0 degree and φ varying from -90 degree to +90 degree with a step of 0.5 degree. RCS variations for a given frequency of 5GHz for horizontal and vertical polarizations for Θ=0 degree and φ varying from -60 degree to +60 degree with a step of 0.5 degree is shown in Figure 4. Flat panel RAS sample shows mono-static RCS reduction of 000dB and 000dB at normal incident angle for horizontal (HH) polarization and vertical (VV) polarizations, respectively. It can also be seen that the flat panel RAS shows the mono-static RCS reduction even at other incident angle (shown from -60 degree to +60 degree). Result shows that the proposed RAS is good performance even at oblique angles as well.

Next, the effect of rivet installation on these flat panel RAS is carried out using three different rivet configurations (refer Section 2 for further details). Figure 4 (a) shows the reflection coefficients of different RAS with and without rivets (with rivet surface area of 9.8%) at 5GHz for HH polarization. It can be seen that the different rivet locations are leading to different mono-static RCS with varying frequencies. For example, RAS C1 shows maximum of 00dB degradation of mono-static RCS at 5GHz of frequency and Θ=0 and φ=0, as shown in Figure 4(a). RAS C2 shows 00dB degradation and RAS C3 shows 00dB degradation compared to RAS (with rivet) for a given case of rivet surface area of 9.8%. The mono-static RCS performance of all the rivet configurations at oblique angle is also shown in Figure 4 at 5GHz of frequency.

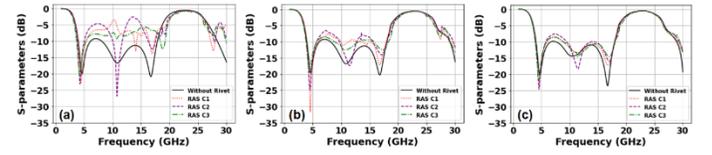

**FIGURE 3:** S-PARAMETERS OF VARYING THE PERCENTAGE OF RIVETS HEADS AREA TO RAS SURFACE AREA OF (a) 9.8%, (b) 4.4%, AND (c) 2.5% FOR DIFFERENT RAS CONFIGURATIONS

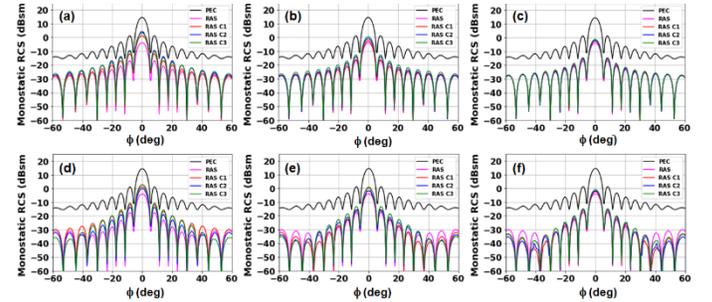

**FIGURE 4:** MONO-STATIC RCS OF VARYING THE PERCENTAGE OF RIVET HEADS AREA TO RAS SURFACE AREA OF (a) 9.8%, (b) 4.4%, AND (c) 2.5% FOR HH POLARIZATION AND (d) 9.8%, (e) 4.4%, AND (f) 2.5% FOR VV POLARIZATION FOR DIFFERENT RAS CONFIGURATIONS

Next, three cases with varying the percentage of rivets heads area to RAS surface area are studied, i.e. (a) 9.8%, (b) 4.4%, and (c) 2.5%. Mono-static RCS of all these configurations are shown in Figures 4(a), 4(b), and 4(c) for rivet surface area of 9.8%, 4.4%, and 2.5%, respectively at 5GHz of frequency and for HH polarization. Figures 4(d), 4(e), and 4(f) show the performance at VV polarization. Result shows that with reduction in rivet surface area to RAS surface area, the degradation in mono-static RCS reduces. Rivet surface area of 4.4% shows reflection coefficient degradation of maximum 00dB and rivet surface area of 2.5% shows maximum of 00dB (except null regions) at normal incident angle for VV polarization at 5GHz of frequency, as shown in Figures 4(b) and 4(c), respectively.

### 3.3 Effect of rivet configurations on different RAS air intake

In this section, the RCS performance of rivets is evaluated on different circular air intake cavities, i.e. metallic intake, RAS intake without rivets, and nine cases of rivets with RAS (refer section 2 for further details). Duct of diameter 0.5m, 1.0m, 2.0m, 4.0m, 6.0m, and 8.0m are considered in the present study. Simulation is carried out from 1 to 18GHz of frequencies with a step size of 1GHz for both HH and VV polarizations for Θ=0 degree and φ varying from 180 degree to 240 degree with a step of 0.5 degree. Effect of φ, RAS configurations, and rivet configurations are discussed next.

#### 3.3.1 Effect of varying angle

Mono-static RCS with varying ϕ varying from 180 degree to 240 degree with a step of 0.5 degree for duct length of 1.0m at 10GHz is shown in Figure 5. Figures 5(a), 5(b), and 5(c) shows the variation of mono-static RCS with varying ϕ for the percentage of rivets heads area to RAS surface area of (a) 9.8%, (b) 4.4%, and (c) 2.5% for HH polarization, respectively. It can be seen that with larger rivet surface area (i.e. 9.8%) shows significant RCS degradation, especially for RAS C1 configuration, as shown in Figure 5(a). It can also be seen that the reduction in mono-static RCS of RAS configuration is less from ϕ 180 degree to 210 degrees. However, the RAS performance increases with further increase in ϕ. Further reduction in rivet surface area leads to improvement in mono-static RCS, as can be seen in Figures 5(b) and 5(c) for rivet surface area of 4.4% and 2.5% for HH polarization, respectively. Similar performance is observed in the VV polarizations, as shown in Figures 5(d), 5(e), and 5(f) for the percentage of rivets heads area to RAS surface area of (a) 9.8%, (b) 4.4%, and (c) 2.5%, respectively.

Further, the mono-static RCS of a longer duct of length 4.0 meter is shown in Figure 6. It can be seen from the Figure 6 that the same RAS configuration given better RCS performance for a longer duct. However, duct with larger rivet surface area (i.e. 9.8%) RAS shows significant RCS degradation, especially for RAS C1 configuration, as shown in Figures 5(a) and 5(d) for HH and VV polarizations, respectively. It has been observed that with an increase in duct length leads to improvement in RCS performance for a given RAS configuration. The impact of duct length is shown in the next section.

### 3.3.2 Effect of duct length

In order to find out the impact of duct length on the mono-static RCS performance, median RCS is plotted with varying ϕ at 10GHz, as shown in Figure 7. Median RCS at a given frequency is also calculated for varying ϕ from 180 degree to 240 degree for a given polarization. Figures 7(a), 7(b), and 7(c) shows the variation of median RCS with varying duct length for the RAS C1, RAS C2, and RAS C3 for HH polarization, respectively. Result shows that a slight increase in median RCS is observed for PEC duct with an increase in duct length. However, duct with all the RAS configurations show improvement in median RCS performance with an increase in duct length. It can also be seen that with larger rivet surface area (i.e. 9.8%) shows significant RCS degradation, especially for RAS C1 configuration, as shown in Figure 7(a). Further reduction in rivet surface area leads to improvement in median RCS, as can be seen in Figures 7(a), 7(b), and 7(c) for rivet surface area of 9.8%, 4.4% and 2.5% for HH polarization, respectively. Similar performance is observed in the VV polarizations, as shown in Figures 7(d), 7(e), and 7(f) for RAS C1, RAS C2, and RAS C3, respectively. Further, the median RCS of duct with varying length at 15.0 GHz is shown in Figure 8. A similar trend of improving RCS performance with an increase in length is observed at 15GHz as well. Also, the RCS improvement with respect to PEC duct is better at 15GHz.

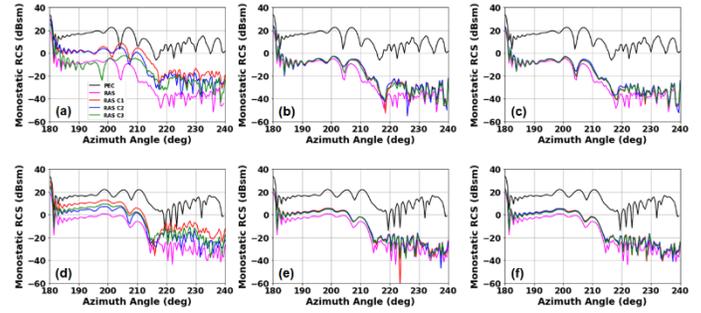

**FIGURE 5:** MONO-STATIC RCS OF VARYING THE PERCENTAGE OF RIVET HEADS AREA TO RAS SURFACE AREA OF (a) 9.8%, (b) 4.4%, AND (c) 2.5% FOR HH POLARIZATION AND (d) 9.8%, (e) 4.4%, AND (f) 2.5% FOR VV POLARIZATION AT FOR DUCT OF 1METER LENGTH AT 10GHz

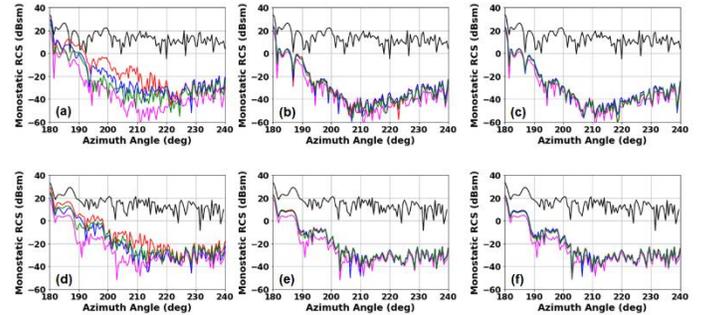

**FIGURE 6:** MONO-STATIC RCS OF VARYING THE PERCENTAGE OF RIVET HEADS AREA TO RAS SURFACE AREA OF (a) 9.8%, (b) 4.4%, AND (c) 2.5% FOR HH POLARIZATION AND (d) 9.8%, (e) 4.4%, AND (f) 2.5% FOR VV POLARIZATION AT FOR DUCT OF 4 METERS LENGTH AT 10GHz

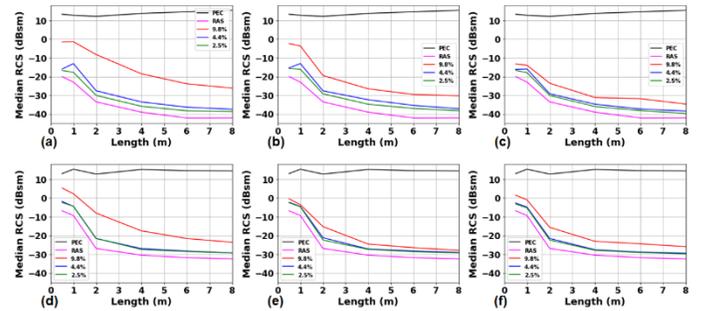

**FIGURE 7:** MEDIAN RCS FOR DIFFERENT LENGTH OF AIR INTAKE DUCT AT A GIVEN FREUENCY OF 10 GHz FOR VARYING RAS CONFIGURATION (a) RAS C1, (b) RAS C2, AND (c) RAS C3 FOR HH POLARIZATION AND (d) RAS C1, (e) RAS C2, AND (f) RAS C3 FOR VV POLARIZATION

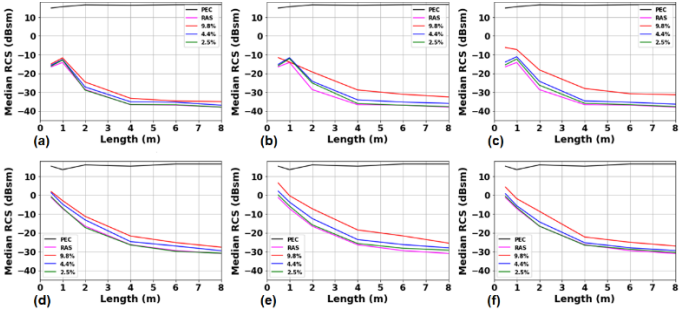

**FIGURE 8:** MEDIAN RCS AT DIFFERENT LENGTH OF AIR INTAKE DUCT FOR A FREQUENCY AT 15 GHz VARYING RAS CONFIGURATION (a) RAS C1, (b) RAS C2, AND (c) RAS C3 FOR HH POLARIZATION AND (d) RAS C1, (e) RAS C2, AND (f) RAS C3 FOR VV POLARIZATION

### 3.4 Effect of rivet configuration on RAS air intake installed on an aircraft

In this section, mono-static RCS performance of an open-source aircraft CAD model is studied. The performance of RAS air intake with and without rivets is evaluated and results are compared with aircraft with non-RAS (i.e. PEC) air intake. The high frequency EM simulations using shooting and bouncing rays are used for cavity study with 20 numbers of bounces with 5 ray densities [19]. Simulation is carried out from 1 to 18GHz of frequencies with a step size of 1GHz for both horizontal and vertical polarizations for Θ=0 degree and ϕ varying from 180 degree to 270 degree with a step of 0.5 degree. ϕ=180 degree corresponds to the nose of the aircraft and ϕ=270 degree is correspond to the broad side of the aircraft. Normalized mono-static RCS of an open-source aircraft with varying ϕ from 180 to 270 degree at (a) 5GHz, (b) 10GHz, and (c) 15GHz for HH polarization is shown in Figures 9(a), 9(b), and 9(c), respectively. It can be seen that all the rivet configurations show almost similar mono-static RCS when installed on RAS air intake duct of aircraft at different frequencies. It can also be seen that the RAS air intake shows mono-static RCS reduction from ϕ=180 degree to 240 degree. Afterwards the RCS reduction of aircraft is not visible. This is due to the face that the CAD model considered in the present study shows higher mono-static RCS at broad side. Also, it can be seen that the mono-static RCS reduction from ϕ=180 degree to 240 degree is around 10dB compared to PEC air intake duct when installed on aircraft. However, the mono-static RCS reduction for circular cavity is very high compared to the aircraft air intake duct. This is mainly due to the fact that the other components/systems of aircraft is also contributing to the overall RCS and hence, it lead to under performance of the RAS when installed on the aircraft. This performance may be further improved with proper selection/design of aircraft's external shape. The external shape modification is not the present scope of this work. In the previous section, it is observed that the RAS C1 shows the worst performance; hence, RAS C1 is considered for evaluating the median RCS performance with varying frequencies, as shown in Figure 10. Median RCS at a given frequency is calculated for varying ϕ from 180 degree to 240 degree for a given polarization. Result shows median RCS of ~10dBsm across the band of frequencies for PEC air intake duct when installed on aircraft. RAS air intake with and without rivet configuration (RAS C1 is considered) shows Median RCS reduction of approximately 10dB from 4 to 18GHz of frequencies for both the HH and VV polarizations, as shown in Figure 10. It is also to be noted that RAS air intake duct with the entire rivet configuration gives similar Median RCS performance across the frequency bands. It can also be seen that below 4GHz the RAS performance is degrading. This is mainly due to the face the RAS configuration considered in the present study works from 4 to 18GHz of frequencies only.

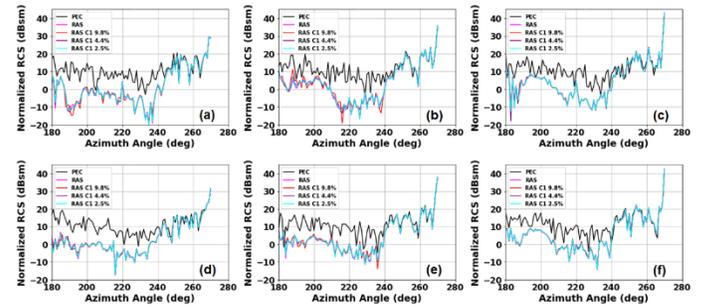

**FIGURE 9:** NORMALIZED RCS OF AN OPEN-SOURCE AIRCRAFT WITH VARYING Φ FROM 180 (NOSE OF THE AIRCRAFT) TO 270 (BROAD SIDE OF THE AIRCRAFT) DEGREE WITH A STEP OF 0.5 DEGREE AT (a) 5GHz, (b) 10GHz, AND (c) 15GHz FOR HH POLARIZATION AND (d) 5GHz, (e) 10GHz, AND (f) 15GHz FOR VV POLARIZATION

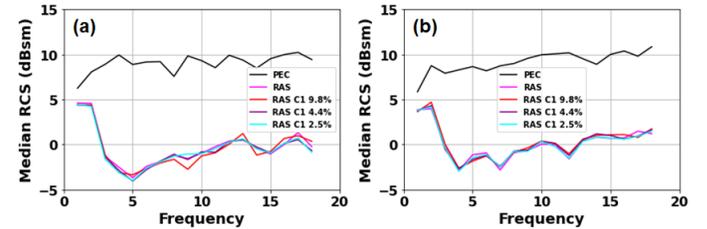

**FIGURE 10:** NORMALIZED MEDIAN RCS OF AN OPEN-SOURCE AIRCRAFT FROM 1-18GHz OF FREQUENCIES FOR (a) HH AND (b) VV POLARIZATION

### 4. CONCLUDING REMARKS

In this paper, impact of rivets on the RCS performance of RAS is carried out in details. Initially, a Jerusalem cross unit cell RAS is designed which performs from 4 to 18GHz of frequency with ~10 dB of RCS reduction. Subsequently, the performance of flat panel RAS with various rivet configurations are studied and its impact on RCS performance is evaluated. Also, circular duct of different length is studied and impact of rivets on the RCS performance on different cavity is evaluated at different frequencies and angles. Result shows that larger rivet surface area will lead to degradation in RCS performance as compared to the RAS duct without rivets. Also, the results show that the location of rivets on RAS may have impact on the overall RCS performance. Lastly, an open-source aircraft CAD model is considered and the performance of RAS air intake with and without rivets is evaluated. It has been observed that for the open-source aircraft CAD model all the RAS configurations give almost similar RCS performance. It can be seen that the RAS shows almost 10 dB RCS reduction from 4 to 18 GHz of frequencies when implemented on the open-source aircraft CAD model. The impact of rivets is negligible in the present aircraft model. However, a low RCS aircraft model, the impact of rivets may be significant. A further study on different shape and size of rivets and its pattern installed on air intake duct is open research area. Also, the impact of rivets on a low observable aircraft will be studied in future and its effect will be reported.

**ACKNOWLEDGEMENTS**

Authors would like to thank Shri. Y Dilip, Director, Aeronautical Development Establishment, Mr. Manjunath S M, Technology Director and Mr. Diptiman Biswas, Group Director for their support during the research work carried out at ADE, DRDO.